\documentclass{aa}
\usepackage{graphics}

\def\teff{${\rm T}_{\rm eff}$}
\def\logg{$\log g$}
\def\met{[M/H]}
\def\vmicro{$V_{\rm micro}$}
\def\abun{\{$\alpha_i$\}}
\def\extinct{A($\lambda)$}
\def\aapix{\AA$/{\rm pix}^{-1}$}

\def\yv{{\bf y}}

\def\xv{{\bf x}}
\def\pv{{\bf p}}
\def\qv{{\bf q}}
\def\wv{{\bf w}}

\begin{document}

   \thesaurus{ 06 \\
		(03.13.2 
		 03.13.4 
		 04.19.1 
		 08.08.1 
		 08.06.3 
		 10.19.2 
		)}
   \title{Stellar parameters from very low resolution spectra and
medium band filters}
   \subtitle{ \teff, \logg\ and \met\ using neural networks}
   \titlerunning{stellar parameters at very low spectral resolution}
   \author{C.A.L.\ Bailer-Jones
	}
   \authorrunning{Bailer-Jones}
   \institute{Max-Planck-Institut f\"ur Astronomie, K\"onigstuhl 17,
              D-69117 Heidelberg, Germany. email: calj@mpia-hd.mpg.de}

   \date{Submitted 16 November 1999; Accepted 18 February 2000}

   \maketitle

\begin{abstract}
Large scale, deep survey missions such as GAIA will collect enormous
amounts of data on a significant fraction of the stellar content of
our Galaxy.  These missions will require a careful optimisation of
their observational systems in order to maximise their scientific
return, and will require reliable and automated techniques for
parametrizing the very large number of stars detected.  To address
these two problems, I investigate the precision to which the three
principal stellar parameters (\teff, \logg, \met)\ can be determined
as a function of spectral resolution and signal-to-noise (SNR) ratio,
using a large grid of synthetic spectra. The parametrization technique
is a neural network, which is shown to provide an accurate
three-dimensional physical parametrization of stellar spectra across a
wide range of parameters.  It is found that even at low resolution
(50--100\,\AA\ FWHM) and SNR (5--10 per resolution element), \teff\
and \met\ can be determined to 1\% and 0.2 dex respectively across a
large range of temperatures (4000--30\,000\,K) and metallicities
($-$3.0 to $+$1.0 dex), and that \logg\ is measurable to $\pm 0.2$ dex
for stars earlier than solar. The accuracy of the results is probably
limited by the finite parameter sampling of the data grid.  The
ability of medium band filter systems (with 10--15 filters) for
determining stellar parameters is also investigated.  Although easier
to implement in a unpointed survey, it is found that they are only
competitive at higher SNRs ($\geq 50$).


\end{abstract}

\section{Background and Objectives}

An understanding of the origin, properties and evolution of our Galaxy
requires a careful census of its constituents, in particular its
stellar members.  Of special importance are the intrinsic physical
properties of these stars. The fundamental properties are mass, age
and abundances, as these determine a star's history and future
development. However, ages are not observable, and masses can only be
directly obtained from some multiple systems.  Thus we must indirectly
gain this information via the stellar spectrum, and a number of
atmospheric parameters have been defined for this purpose. The main
ones are the effective temperature, \teff, the surface gravity, \logg,
and the metallicity, \met.  To these can also be added the alpha
abundances, \abun\ (which measure the devations away from the
`standard' abundance ratios), the photospheric microturbulence
velocity, \vmicro, and the extinction by the interstellar medium,
\extinct (although not intrinsic to the star, it is necessary for
determining its luminosity). Masses and ages can then be determined
from stellar structure and evolution models and with calibration via
binary systems. It is important to realise that this modelling is
complex, and a number of assumptions have to be made. There is,
therefore, a limit to the precision with which we can determine
physical properties.

Historically, spectroscopic parameters have been measured indirectly
through the MK classification system (Morgan et
al.\ \cite{morgan_43a}) or via colour-magnitude and colour-colour
diagrams. In the MK system, the two parameters {\it spectral type} and
{\it luminosity class} act as proxies for \teff\ and \logg. Originally
a qualitative system relying on a visual match between observed
spectra and a system of standards, much progress has been made in
quantifying it with automated techniques (e.g.\ Weaver \&
Torres-Dodgen \cite{weaver_97a}; Bailer-Jones et al.\
\cite{bailerjones_98a}). The most commonly used classification
techniques have been neural networks and $\chi^2$ matching to
templates (or more generally, minimum distance methods).  A summary of
recent progress in this area is given by von Hippel \& Bailer-Jones
(\cite{vonhippel_00a}).

Despite this focus on the MK system, it is not well suited to
classifying data from the deep surveys which will be central to the
future development of Galactic astrophysics. This is for a number of
reasons, but in particular because it lacks a measure of
metallicity. Although MK does make allowance for various `peculiar'
stars, these are defined as exceptions, and the notation is not suited
to a statistical, quantifiable analysis.  This is problematic given
the significance of metal poor halo stars in a deep survey.  There is
also now no good reason why we should not determine physical
parameters directly from the observational data.

Some attempts have been made to determine the physical parameters
of real spectra directly by training neural networks on synthetic
spectra. Gulati et al.\ (\cite{gulati_97a}) used this approach
to determine the effective temperatures of ten solar metallicity G and
K dwarfs. Taking the ``true'' effective temperature of these stars as
those given by Gray \& Corbally \cite{gray_94a}, they found a mean
``error'' in the network-assigned temperatures of 125\,K.
Bailer-Jones et al.\ (\cite{bailerjones_97a}) determined \teff\ for
over 5000 dwarfs and giants in the range B5--K5, and also showed
evidence of sensitivity of the parametrization models to metallicity.

The accuracy with which physical parameters can be determined from a
stellar spectrum depends upon, amongst other things, the wavelength
coverage, spectral resolution and signal-to-noise ratio (SNR). From
the point of view of designing a stellar survey project it is
essential to know how well the stellar parameters can be determined
for a given set of these observational parameters. Moreover, given
that there is always a limit to the collecting area and integration
time available, there is always a trade-off between spectral
resolution, sensitivity and sky coverage.  

The goal of this paper is to determine the accuracy with which
physical stellar parameters can be determined from spectroscopic data
at a range of SNRs and resolutions which could realistically be
achieved in a {\it deep} survey mission. This specification rules out
high resolution spectra. The parametrization work has been carried out
using neural networks (Sect.~\ref{networks}) because they have been
shown to be one of the best approaches for this kind of work. This is
not to presuppose, however, that some other approach may not
ultimately be better.  The simulations have been made using a large
database of synthetic spectra generated from Kurucz atmospheric models
(Sect.~\ref{synspec}). While these spectra do not show the full
range of variation in real stellar spectra, they are adequate for a
realistic demonstration of what is possible as a function of SNR and
resolution. The results are presented in Sect.~\ref{results} and
summarised and discussed in Sect.~\ref{discussion}.  Finally, the
requirements for a complete survey-oriented classification system are
given in Sect.~\ref{system}.

\section{The GAIA Galactic Survey Mission}\label{gaia}

The simulations in this paper were partially inspired by the need to
produce an optimal photometric/spectroscopic system for the GAIA
Galactic survey mission. GAIA is a candidate for the ESA cornerstone 5
mission for launch in 2009 (ESA, in preparation). It is primarily an
astrometric mission with a precision of a few microarcseconds, and
will survey the entire sky down to V=20, thus observing c.\ 10$^9$
stars in our Galaxy. Radial velocities will be obtained on board down
to V=17.5, thus providing a 6D phase space survey (three spatial and
three velocity co-ordinates) for stars brighter than this limit. A
survey of this size will have a profound impact on Galactic
astrophysics, but to achieve this it is essential that the physical
characteristics of the target objects are measured and correlated with
their spatial and kinematic properties. As GAIA is a continuously
scanning satellite, a fixed total amount of integration time is
available for each object, so there is a trade-off between resolution,
signal-to-noise ratio and wavelength coverage.  For various reasons,
the current GAIA design does not include a spectrograph (other than a
1.5\,\AA\ resolution region between 8470 and 8700\,\AA\ intended for
radial velocity measurements), but instead will image all objects in
several medium and broad band filters (Table~\ref{filttab}).  Three
filter systems are shown: the system nominally selected for the
mission plus two alternatives.  The profiles of the two alternatives
are represented as Gaussians in this paper.  The ability of these
filter system to determine stellar parameters will be compared with
that for spectra of various resolutions.

\begin{table}[t]
\begin{center}
\caption{Three multiband filter systems proposed for the GAIA mission.
All profiles are symmetric about the central wavelength, $\lambda_c$,
and have a FWHM of $\Delta\lambda$. The profiles of the filters in the
Asiago and modified Str\"omvil systems (F.\ Favata 1999, private
communication) are defined as Gaussians (although note that the former
is only an approximation to the original Asiago system in Munari
\cite{munari_99a}).  The filters of the selected GAIA system (ESA, in
preparation) have flatter tops and steeper sides than Gaussians, and
have defined relative peak transmissions, T. There is some (intended)
redundancy within each filter system.}
\label{filttab}
\begin{tabular}{cr|cr|crc}\hline
\multicolumn{2}{c}{Asiago} & \multicolumn{2}{|c|}{mod Str\"omvil} & \multicolumn{3}{c}{GAIA} \\ \hline 
$\lambda_c$\,/\,\AA & $\Delta\lambda$\,/\,\AA  &  $\lambda_c$\,/\,\AA & $\Delta\lambda$\,/\,\AA  &  $\lambda_c$\,/\,\AA & $\Delta\lambda$\,/\,\AA & T \\ \hline
 3000 &  1410 &  3450 &   400 &  3260 &   820 &  0.92 \\
 3860 &   190 &  3800 &   300 &  3750 &  1460 &  0.96 \\
 4090 &   170 &  4050 &   200 &  4050 &   600 &  0.90 \\
 4300 &   120 &  4450 &  1100 &  4645 &   450 &  0.86 \\
 4800 &  1500 &  4600 &   200 &  5075 &   270 &  0.78 \\
 5270 &    80 &  5150 &   200 &  5250 &  2070 &  0.97 \\
 5310 &   170 &  5450 &   200 &  5700 &   900 &  0.93 \\ 
 6300 &  1500 &  5500 &  1000 &  6560 &   240 &  0.72 \\
 7920 &  1720 &  6500 &  1000 &  6740 &  1160 &  0.94 \\
 9640 &  1700 &  6560 &   200 &  7330 &  1850 &  0.97 \\
      &       &  7500 &  1000 &  7470 &   280 &  0.79 \\
      &       &  8000 &   400 &  7775 &   310 &  0.81 \\
      &       &  8500 &  1000 &  8160 &   480 &  0.87 \\
      &       &  8700 &   300 &  8940 &   480 &  0.97 \\ 
      &       &  9380 &   200 &       &       &       \\ \hline
\end{tabular}
\end{center}
\end{table}

\section{The network model}\label{networks}

A neural network is an algorithm which performs a non-linear
parametrized mapping between an input vector, $\xv$, and an output
vector, $\yv$.  (The term `neural' is misleading: although originally
developed to be very simplified models of brain function, many neural
networks have nothing to do with brain research and are better
described in purely mathematical terms.)  The network used in this
paper is a feedforward multilayer perceptron with two `hidden
layers'. These hidden layers form non-linear combinations of their
inputs. The output from the first hidden layer is the vector $\pv$,
the elements of which are given by
\[p_j = \tanh \left( \sum_i w_{i,j} x_i \right) \]
These values are then passed through a second hidden layer which performs
a similar mapping, the output from that layer being the vector $\qv$
\[q_k = \tanh \left( \sum_j w_{j,k} p_j \right) \]
The output from the network, $\yv$, is then the weighted sum of these
\[y_l =  \sum_k w_{k,l} q_k\]
The $\tanh$ function provides the non-linear capability of the
network, and the weights, $\wv$, are its free parameters.  The model
is supervised, which means that in order for it to give the required
input--output mapping it must be trained on a set of representative
data patterns. These are inputs (stellar spectra) for which the true
{\em target} outputs (stellar parameters) are known. The training is a
numerical least-squares minimisation: Starting with random values for
the weights, a set of spectra are fed through the network and the
error in the actual outputs with respect to the desired (target)
outputs calculated. The gradient of this error with respect to each of
the N weights is then used to iteratively perturb the weights towards
a minimum of the error function.  Thus the training is a minimisation
problem in an N-dimensional space, and the resulting input--output
mapping can be regarded as a non-linear interpolation of the training
data.  Once the network has been trained the weights are fixed and the
network used to obtain physical stellar parameters for new spectra.

The results in this paper use a network code written by the author
consisting of five and ten hidden nodes in the first and second hidden
layers respectively. The complexity of the network is determined by
the number of hidden nodes and layers.  While networks with a
single hidden layer can provide non-linear mappings, experience has
shown that a second hidden layer can lead to considerable improvement
in performance (Bailer-Jones et al.\ \cite{bailerjones_98a}).  This
has been confirmed with the data in this paper. Significant further
improvement is not expected through the addition of more hidden
nodes/layers. The network has three outputs, one for each of the
parameters \teff, \logg\ and \met.  The error which is minimised is
the commonly-used sum-of-squares error (the sum being over all
training patterns and outputs), except that the error contribution
from each output is weighted by a factor related to the precision with
which that parameter can be determined.

I stress that a neural network is not fundamentally different from
many other parameter fitting algorithms. Its strengths are that it has
a fast and straight-forward training algorithm, can map arbitrarily
complex functions (given sufficient data to determine the function),
and can be parallelised in software or hardware to achieve
considerable increases in speed. One of the common criticisms of
neural networks is that it is difficult to interpret their weights and
get an idea of exactly {\it how} they achieve their results.  While
this is essentially true, part of this difficulty stems from the fact
that the models are problem-independent: they are purely mathematical
models that do not explicitly take into account the physics of
the problem.  Moreover, in order to fully understand the model it
would be necessary to simplify it, and this in turn would reduce its
performance. This ``interpretability--complexity''
trade-off is inherent to almost any type of heuristic model.

\section{Synthetic spectra}\label{synspec}

A large grid of synthetic spectra have been generated using Kurucz
atmospheric models (Kurucz \cite{kurucz_92a}) and the synthetic
spectral generation program of Gray (Gray \& Corbally
\cite{gray_94a}).  The parameter grid consists of 36 \teff\ values
between 4000\,K and 30\,000\,K (step sizes between 250\,K and
5000\,K), 7 values of \logg\ between 2.0 and 5.0 dex (in 0.5 steps)
and 15 values of \met\ between $-$3.0 and $+$1.0 dex (step sizes
between 0.1 and 0.5). The microturbulence velocity was fixed at
2.0\,kms$^{-1}$. This yielded an (almost complete) grid of 3537
atmospheric models. Contiguous spectra were calculated between 3000
and 10\,000\,\AA\ in 0.05\,\AA\ steps with a line list of over
900\,000 atomic and molecular lines. The resolution, $r$, of these
spectra was then degraded to 25, 50, 100, 200 and 400 \AA\ FWHM by
Gaussian convolution. (Each resolution element is sampled by two
pixels, so these resolutions correspond to 560, 280, 140, 70 and 35
inputs to the network respectively.)  These resolutions are
considerably lower then the 1--5\,\AA\ generally used for MK
classification.  The spectra were also combined with the transmission
curves of the filters (Table~\ref{filttab}) to produce three sets of
filter fluxes.  Poisson noise was added to all data sets to simulate
signal-to-noise ratios of 5, 10, 20, 50 and 1000 per resolution
element. The result is 40 sets of 3537 absolute spectral energy
distributions at each combination of resolution and SNR.  The absolute
flux information is retained.

It is noted that Kurucz models do not produce highly accurate spectra
for all types of stars. This is particularly true at low \teff\ as
they exclude water opacity (and there are no H$_2$O lines in the line
lists).  For this reason spectra have not been calculated below
4000\,K.  Furthermore, the models lack chromospheres and so do not
reproduce features such as emission in the cores of the Ca{\small II}
H\,\&\,K absorption lines. For the present investigation, however, it
is not necessary to have highly accurate individual spectra, but
spectra which reflect differences of the appropriate scale and
complexity.

\section{Spectral parametrization results}\label{results}

As the neural network is a parameter fitting algorithm, it is
essential that its performance is evaluated on an independent set of
data from that on which it is trained. For this purpose, each of the
40 data sets was randomly split into two halves and one used for
training (1760 spectra) and the other for testing (1759 spectra).
$\log_{\rm 10}$\teff\ (rather than \teff)\ is used as a target in the
networks to reduce the dynamic range of this parameter and give a
better representation of the uncertainties. The input and output
parameters are scaled to have zero mean and unit standard deviation to
prevent `saturation' of the network during training.

For each data set a {\em committee} of three identical networks was
trained from different initial random weights. The resultant parameter
for any star is then the average from the three networks. This helps
to reduce the effects of imperfect training convergence. Each network
was trained with a conjugate gradient algorithm for 10\,000 iterations
and used weight decay regularisation to avoid overtraining. More
training did not reduce the error further.  The longest training time
(for the largest input vector) was about one day on a Sun SPARC
Enterprise (no parallelisation of the code). The time to parametrize
was of order $10^{-3}$ seconds per spectrum.

The precision to which physical parameters can be determined from a
stellar spectrum depends not only on the SNR and resolution, but also
on the type of star. For example, it is more difficult to determine
the metallicity of hot stars on account of the almost complete absence
of metal lines. Therefore, I summarise the performance of each data
set for three different temperature ranges (for all \logg\ and \met):
\begin{enumerate}
\item{\teff\ $<$ 5800\,K (stars later than solar -- 408 spectra in the test subset)}
\item{5800 $<$ \teff\ $<$ 10\,000 (A and F stars -- 888 spectra in the test subset)}
\item{\teff\ $>$ 10\,000\,K (O and B stars -- 463 spectra in the test subset)}
\end{enumerate}
The error measure I used is the average absolute error, $\epsilon$, of
each parameter, i.e.\ the absolute difference between the network
output and the target value averaged over all stars in the test subset
for that temperature range.  This error is more robust than the
often-used RMS error because it is less distorted by outliers and more
characteristic of the majority of the error distribution. For a
Gaussian distribution $1\sigma=1.25\epsilon$, although some of the
error distributions deviate significantly from Gaussian.


\begin{figure*}[t]
\resizebox{\hsize}{!}{\includegraphics{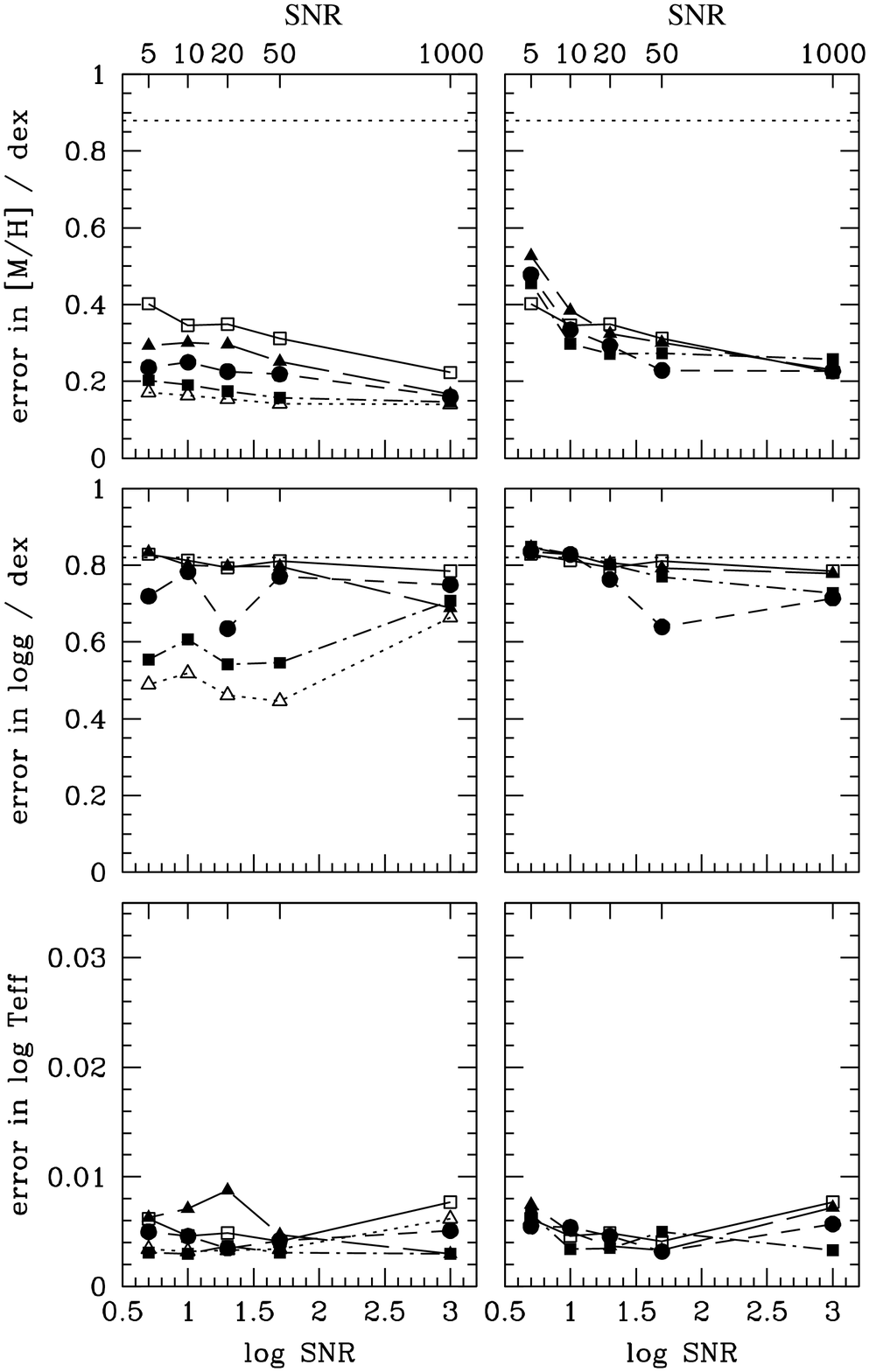}}
\caption{\teff\ $<$ 5800\,K.  Error in the determination of physical
parameters as a function of SNR for spectra at different resolutions
(left column) and for three sets of filters (right column). The
different resolutions shown in the left column are 25\,\AA\ (open
triangles, dotted line), 50\,\AA\ (filled squares, dot-dash line),
100\,\AA\ (filled circles, short dashed line), 200\,\AA\ (filled
triangles, long dashed line) and 400\,\AA\ (open squares, solid line).
The three filter systems in the right column are Asiago (filled
circles, short dashed line), modified Str\"omvil (filled squares,
dot-dash line) and GAIA (filled triangles, long dashed line), and the
$r$\,=\,400\,\AA\ results are shown again for comparison (open
squares, solid line).  For all plots the vertical axis is the mean
absolute error, $\epsilon$, across all spectra in the test subset in
this temperature range. Note that the fractional error in \teff\ is
equal to 2.3 times the error in $\log_{\rm 10}$\teff. The horizontal
dotted lines on the \logg\ and \met\ plots are the performances of
random (untrained) networks. This has a small dependence on the
resolution (number of inputs), so the minimum values are shown.  The
corresponding value for \teff\ is $\epsilon=0.13$. The results are
tabulated in Tables~\ref{restab1}--\ref{restab3}. }
\label{results1}
\end{figure*}

\begin{figure*}[t]
\resizebox{\hsize}{!}{\includegraphics{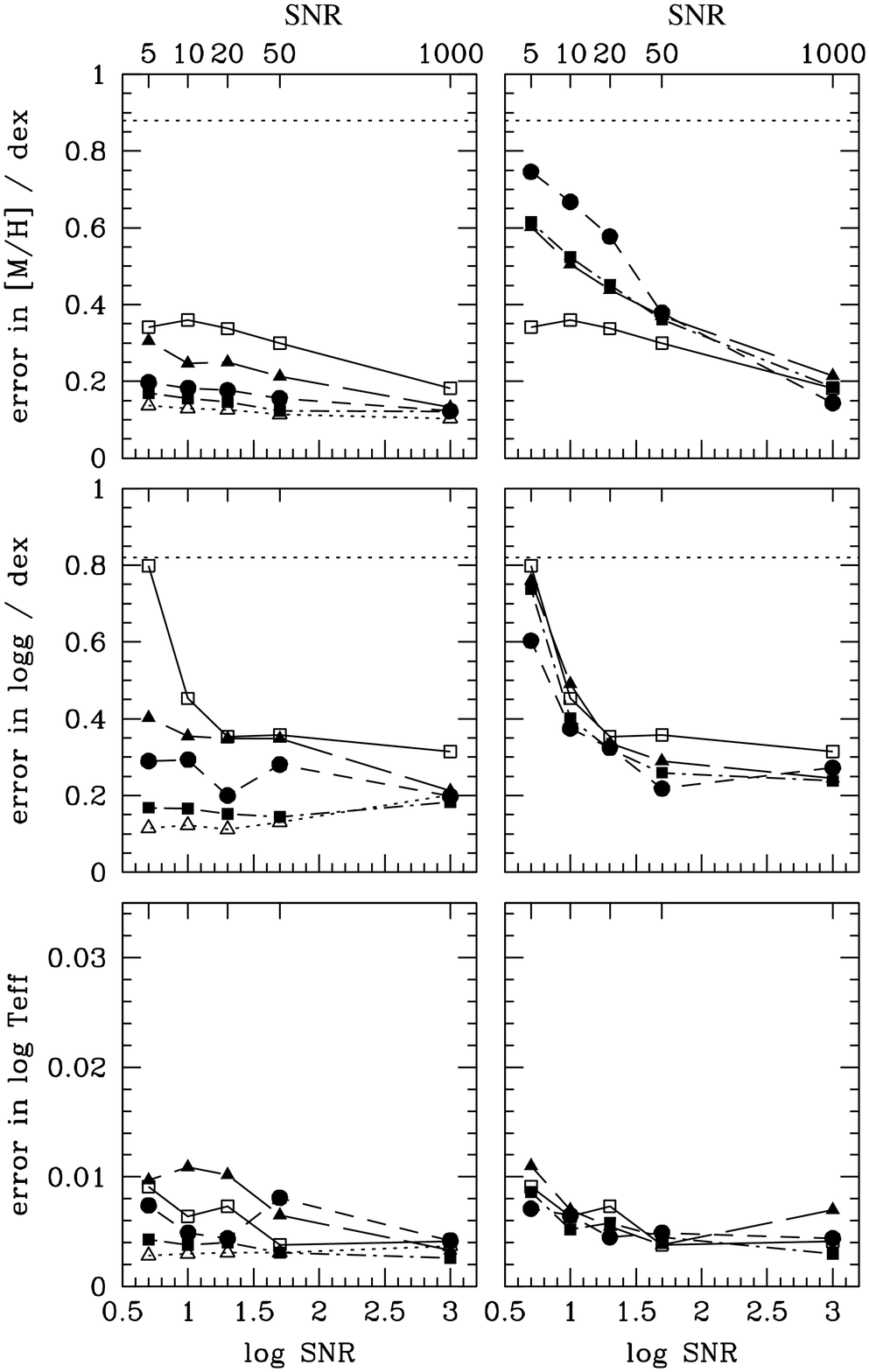}}
\caption{Same as Fig~\ref{results1} but for 5800 $<$ \teff\ $<$ 10\,000.}
\label{results2}
\end{figure*}

\begin{figure*}[t]
\resizebox{\hsize}{!}{\includegraphics{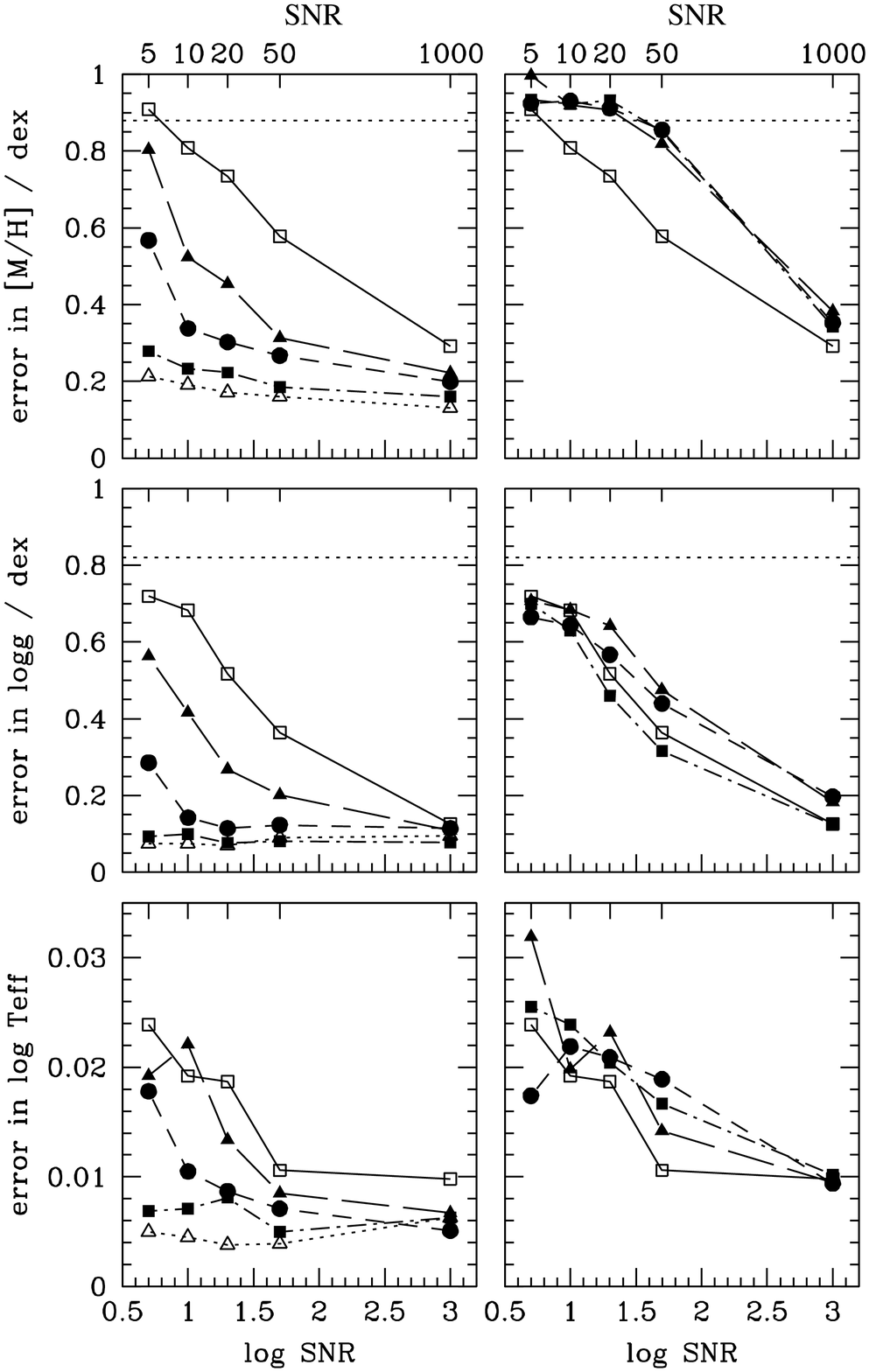}}
\caption{Same as Fig~\ref{results1} but for \teff\ $>$ 10\,000\,K.}
\label{results3}
\end{figure*}

\begin{table}
\begin{center}
\caption{\met\ accuracy. Tabulation of the results in
Figs~\ref{results1}--\ref{results3}.  The resolution is in \AA, except
for the three filter systems which are denoted by their names. SNR is
the signal-to-noise ratio (per resolution element in the case of the
spectra). $\epsilon_1$, $\epsilon_2$ and $\epsilon_3$ are the mean
absolute errors for the three temperature ranges $<$\,5800,
5800--10\,000 and $>$\,10\,000\,K respectively. $\epsilon_{\rm all}$
is the error across all temperatures (4000--30\,000\,K).}
\label{restab1}
\begin{tabular}{rrcccc} \hline
resolution  &   SNR  &  $\epsilon_1$ & $\epsilon_2$ & $\epsilon_3$ &  $\epsilon_{\rm all}$ \\ \hline
Asiago  &  1000  &  0.227	 &  0.144  &  0.353  &  0.218  \\
     &    50  &  0.229	 &  0.379  &  0.855  &  0.464  \\
     &    20  &  0.293	 &  0.577  &  0.911  &  0.593  \\
     &    10  &  0.334	 &  0.668  &  0.930  &  0.653  \\
     &     5  &  0.478	 &  0.746  &  0.924  &  0.726  \\ \hline
modified &  1000  &  0.258	 &  0.185  &  0.343  &  0.243  \\
Str\"omvil     &    50  &  0.273	 &  0.362  &  0.852  &  0.465  \\
     &    20  &  0.272	 &  0.451  &  0.932  &  0.530  \\
     &    10  &  0.296	 &  0.523  &  0.923  &  0.570  \\
     &     5  &  0.455	 &  0.616  &  0.933  &  0.657  \\ \hline
GAIA &  1000  &  0.230	 &  0.215  &  0.382  &  0.261  \\
     &    50  &  0.301	 &  0.370  &  0.818  &  0.468  \\
     &    20  &  0.324	 &  0.438  &  0.907  &  0.530  \\
     &    10  &  0.385	 &  0.506  &  0.920  &  0.582  \\
     &     5  &  0.528	 &  0.603  &  0.996  &  0.685  \\ \hline
400  &  1000  &  0.223	 &  0.182  &  0.292  &  0.220  \\
     &    50  &  0.312	 &  0.300  &  0.578  &  0.376  \\
     &    20  &  0.349	 &  0.337  &  0.735  &  0.445  \\
     &    10  &  0.346	 &  0.359  &  0.808  &  0.474  \\
     &     5  &  0.402	 &  0.341  &  0.908  &  0.505  \\ \hline
200  &  1000  &  0.167	 &  0.132  &  0.222  &  0.164  \\
     &    50  &  0.252	 &  0.213  &  0.313  &  0.248  \\
     &    20  &  0.296	 &  0.251  &  0.454  &  0.315  \\
     &    10  &  0.301	 &  0.247  &  0.524  &  0.332  \\
     &     5  &  0.294	 &  0.305  &  0.803  &  0.434  \\ \hline
100  &  1000  &  0.160	 &  0.123  &  0.199  &  0.151  \\
     &    50  &  0.219	 &  0.156  &  0.267  &  0.200  \\
     &    20  &  0.226	 &  0.177  &  0.302  &  0.221  \\
     &    10  &  0.250	 &  0.182  &  0.338  &  0.239  \\
     &     5  &  0.236	 &  0.198  &  0.568  &  0.304  \\ \hline
 50  &  1000  &  0.147	 &  0.121  &  0.161  &  0.138  \\
     &    50  &  0.158	 &  0.123  &  0.186  &  0.147  \\
     &    20  &  0.174	 &  0.146  &  0.223  &  0.173  \\
     &    10  &  0.191	 &  0.155  &  0.232  &  0.184  \\
     &     5  &  0.203	 &  0.169  &  0.279  &  0.206  \\ \hline
 25  &  1000  &  0.140	 &  0.103  &  0.132  &  0.119  \\
     &    50  &  0.141	 &  0.113  &  0.160  &  0.132  \\
     &    20  &  0.154	 &  0.126  &  0.172  &  0.145  \\
     &    10  &  0.164	 &  0.129  &  0.191  &  0.154  \\
     &     5  &  0.170	 &  0.137  &  0.214  &  0.165  \\ \hline
\end{tabular}
\end{center}
\end{table}

\begin{table}
\begin{center}
\caption{\logg\ accuracy. See Table~\ref{restab1} for details}
\label{restab2}
\begin{tabular}{rrcccc} \hline
resolution  &   SNR  &  $\epsilon_1$ & $\epsilon_2$ & $\epsilon_3$ &  $\epsilon_{\rm all}$ \\ \hline
Asiago & 1000 & 0.714	 &  0.272  &  0.197  &  0.362  \\
     &    50  &  0.640	 &  0.218  &  0.440  &  0.379  \\
     &    20  &  0.763	 &  0.325  &  0.567  &  0.494  \\
     &    10  &  0.828	 &  0.375  &  0.644  &  0.555  \\
     &     5  &  0.836	 &  0.604  &  0.665  &  0.676  \\ \hline
modified  &  1000  &  0.728	 &  0.238  &  0.125  &  0.330  \\
Str\"omvil  &    50  &  0.770	 &  0.260  &  0.316  &  0.400  \\
     &    20  &  0.801	 &  0.322  &  0.459  &  0.475  \\
     &    10  &  0.829	 &  0.401  &  0.631  &  0.565  \\
     &     5  &  0.849	 &  0.738  &  0.699  &  0.755  \\ \hline
GAIA &  1000  &  0.778	 &  0.246  &  0.183  &  0.361  \\
     &    50  &  0.792	 &  0.290  &  0.476  &  0.461  \\
     &    20  &  0.807	 &  0.336  &  0.643  &  0.530  \\
     &    10  &  0.826	 &  0.491  &  0.684  &  0.623  \\
     &     5  &  0.849	 &  0.760  &  0.707  &  0.768  \\ \hline
400  &  1000  &  0.785	 &  0.315  &  0.126  &  0.374  \\
     &    50  &  0.811	 &  0.357  &  0.364  &  0.465  \\
     &    20  &  0.793	 &  0.353  &  0.517  &  0.498  \\
     &    10  &  0.813	 &  0.453  &  0.683  &  0.597  \\
     &     5  &  0.829   &  0.799  &  0.719  &  0.785  \\ \hline
200  &  1000  &  0.689	 &  0.212  &  0.108  &  0.295  \\
     &    50  &  0.797	 &  0.349  &  0.206  &  0.414  \\
     &    20  &  0.797	 &  0.348  &  0.268  &  0.431  \\
     &    10  &  0.800	 &  0.354  &  0.416  &  0.474  \\
     &     5  &  0.834	 &  0.402  &  0.564  &  0.545  \\ \hline
100  &  1000  &  0.750	 &  0.198  &  0.115  &  0.304  \\
     &    50  &  0.770	 &  0.281  &  0.123  &  0.353  \\
     &    20  &  0.635	 &  0.200  &  0.115  &  0.279  \\
     &    10  &  0.783	 &  0.294  &  0.142  &  0.367  \\
     &     5  &  0.719	 &  0.290  &  0.286  &  0.388  \\ \hline
 50  &  1000  &  0.708	 &  0.183  &  0.078  &  0.277  \\
     &    50  &  0.546	 &  0.144  &  0.081  &  0.221  \\
     &    20  &  0.542	 &  0.152  &  0.077  &  0.223  \\
     &    10  &  0.607	 &  0.166  &  0.100  &  0.251  \\
     &     5  &  0.554	 &  0.168  &  0.093  &  0.238  \\ \hline
 25  &  1000  &  0.665	 &  0.202  &  0.094  &  0.281  \\
     &    50  &  0.446	 &  0.131  &  0.090  &  0.193  \\
     &    20  &  0.462	 &  0.112  &  0.070  &  0.182  \\
     &    10  &  0.520	 &  0.122  &  0.075  &  0.202  \\
     &     5  &  0.489	 &  0.115  &  0.075  &  0.191  \\ \hline
\end{tabular}
\end{center}
\end{table}

\begin{table}
\begin{center}
\caption{\teff\ accuracy. See Table~\ref{restab1} for details}
\label{restab3}
\begin{tabular}{rrcccc} \hline
resolution  &   SNR  &  $\epsilon_1$ & $\epsilon_2$ & $\epsilon_3$ &  $\epsilon_{\rm all}$ \\ \hline
Asiago     &  1000  &  0.0057	 &  0.0044  &  0.0094  &  0.0060  \\
     &    50  &  0.0032	 &  0.0049  &  0.0189  &  0.0081  \\
     &    20  &  0.0046	 &  0.0045  &  0.0209  &  0.0087  \\
     &    10  &  0.0054	 &  0.0065  &  0.0219  &  0.0102  \\
     &     5  &  0.0055	 &  0.0071  &  0.0174  &  0.0093  \\ \hline
modified  &  1000  &  0.0033	 &  0.0030  &  0.0102  &  0.0049  \\
Str\"omvil  &    50  &  0.0050	 &  0.0045  &  0.0167  &  0.0077  \\
     &    20  &  0.0035	 &  0.0058  &  0.0204  &  0.0089  \\
     &    10  &  0.0034	 &  0.0052  &  0.0239  &  0.0095  \\
     &     5  &  0.0066	 &  0.0086  &  0.0255  &  0.0124  \\ \hline
GAIA &  1000  &  0.0072	 &  0.0070  &  0.0095  &  0.0077  \\
     &    50  &  0.0033	 &  0.0038  &  0.0142  &  0.0063  \\
     &    20  &  0.0037	 &  0.0055  &  0.0232  &  0.0096  \\
     &    10  &  0.0050	 &  0.0070  &  0.0198  &  0.0098  \\
     &     5  &  0.0075	 &  0.0110  &  0.0319  &  0.0155  \\ \hline
400  &  1000  &  0.0077	 &  0.0041  &  0.0098  &  0.0064  \\
     &    50  &  0.0041	 &  0.0038  &  0.0106  &  0.0057  \\
     &    20  &  0.0049	 &  0.0073  &  0.0187  &  0.0097  \\
     &    10  &  0.0046	 &  0.0064  &  0.0192  &  0.0093  \\
     &     5  &  0.0062	 &  0.0091  &  0.0239  &  0.0123  \\ \hline
200  &  1000  &  0.0030	 &  0.0033  &  0.0067  &  0.0041  \\
     &    50  &  0.0047	 &  0.0065  &  0.0085  &  0.0066  \\
     &    20  &  0.0088	 &  0.0102  &  0.0134  &  0.0107  \\
     &    10  &  0.0071	 &  0.0109  &  0.0221  &  0.0130  \\
     &     5  &  0.0063	 &  0.0097  &  0.0192  &  0.0114  \\ \hline
100  &  1000  &  0.0051	 &  0.0042  &  0.0051  &  0.0046  \\
     &    50  &  0.0042	 &  0.0081  &  0.0071  &  0.0070  \\
     &    20  &  0.0035	 &  0.0044  &  0.0087  &  0.0053  \\
     &    10  &  0.0046	 &  0.0049  &  0.0105  &  0.0063  \\
     &     5  &  0.0050	 &  0.0074  &  0.0178  &  0.0096  \\ \hline
 50  &  1000  &  0.0030	 &  0.0026  &  0.0063  &  0.0036  \\
     &    50  &  0.0031	 &  0.0031  &  0.0050  &  0.0036  \\
     &    20  &  0.0037	 &  0.0040  &  0.0081  &  0.0050  \\
     &    10  &  0.0030	 &  0.0038  &  0.0071  &  0.0045  \\
     &     5  &  0.0031	 &  0.0043  &  0.0069  &  0.0047  \\ \hline
 25  &  1000  &  0.0062	 &  0.0037  &  0.0063  &  0.0050  \\
     &    50  &  0.0034	 &  0.0031  &  0.0039  &  0.0034  \\
     &    20  &  0.0033	 &  0.0031  &  0.0038  &  0.0033  \\
     &    10  &  0.0032	 &  0.0030  &  0.0045  &  0.0034  \\
     &     5  &  0.0034	 &  0.0028  &  0.0050  &  0.0035  \\ \hline
\end{tabular}
\end{center}
\end{table}

The results of the parametrization process are shown in
Figs~\ref{results1}--\ref{results3} and tabulated in
Tables~\ref{restab1}--\ref{restab3}.  Before interpreting these
results we should consider the limits which the data themselves place
on the performance. First, the network will be unable to produce
errors smaller than the smallest variations in the data set. If, to
take a hypothetical example, the spectra did not change as the
metallicity changed by 1.0 dex, we could not expect the network to
determine \met\ to much better than 0.5 dex. Second, the grid of
atmospheric models represents the physical parameters at a finite
sampling, e.g.\ a constant step size of 0.5 dex for \logg. This
sampling does not in itself limit the precision achievable; it is
perfectly possible for the network to legitimately give an error much
smaller than the sampling because the network is minimising a
continuous error function and not just obtaining the best match
between a spectrum and a set of templates. Nonetheless, the network
input--output mapping is an {\em interpolation} of the training data,
and the more coarsely sampled the parameter grid the harder it is for
the network to get a reliable interpolation.  Consequently, while the
network {\it may} be able to achieve sub-sampling accuracy, we should
not be surprised if it cannot. Thus to avoid over-interpreting these
results we should not compare two errors which are both smaller than
half the sampling level. The {\it average} `half-sampling' values for
\met\ and \logg\ are 0.2 and 0.25 respectively, and for log~\teff\ in
the three temperature ranges (cool, intermediate and hot) are 0.01,
0.01 and 0.03 respectively.  The implication is that, if the network
produces errors smaller than these half-sampling values (as it does),
we cannot know whether the performance is limited by the network model
or by the data themselves. A distinction will only be possible with a
more sensitive and finely sampled grid of atmospheric models.

With the above caveat taken into account, I draw attention to some
interesting features in Figs~\ref{results1}--\ref{results3}.
\begin{enumerate}
\item{Good \teff\ determination is possible with all
resolutions/filter systems and SNRs.  The larger error in \teff\ above
10\,000\,K may be an artifact of the larger half-sampling value in
this region ($\geq$1000\,K).}
\item{Only at high resolution can \logg\ be determined for the coolest
stars and even then the determination is poor relative to the hotter
stars. This is understandable, at least in part, because the \logg\
spectral signature is primarily in the line widths which are only
apparent at high resolution.}
\item{Although the three filter systems differ somewhat, they give
essentially the same performance as each other.}
\item{The filter systems (each with 10--15 input parameters) have
similar \logg\ and \teff\ as the $r$=400\,\AA\ spectra (35 inputs).}
\item{At low SNR, the $r$=400\,\AA\ spectra and the filters give poor
\met\ and very poor \logg\ determination for all three temperature
ranges.}
\item{At high SNR (1000) all resolutions/filter systems appear to be
equally good at determining any of the parameters. 
Differences will probably become apparent with a more sensitive
training grid.}
\item{At higher temperatures the accuracy is more sensitive to SNR than
at lower temperatures.}
\item{Metallicity determination
is more difficult at higher temperatures, especially for the filters and 
low resolution spectra. This is understandable as at high
temperature there are fewer and weaker metal lines which are only
significant at high SNR and/or resolution.}
\item{In most cases there is little difference between the
performances of the $r$=25, 50 and 100\,\AA\ spectra, at least for
this data grid.}
\end{enumerate}

\section{Summary and Discussion}\label{discussion}

The results demonstrate that a fully automated neural network can
accurately determine the three principal physical parameters from
spectroscopic or photometric stellar data, something which has not
previously been demonstrated. Moreover, this work has used spectra of
considerably lower resolution than have been used before in automated
classifiers.  Even at low resolution (50--100\,\AA\ FWHM) and SNR
(5--10 per resolution element), neural networks can yield good
determinations of \teff\ and \met, and even for \logg\ for stars
earlier than solar. Still lower resolutions permit good results
provided the SNR is high enough ($\geq 50$).  That good \teff\ can be
achieved even at low resolution and SNR is perhaps not surprising when
we consider that the spectra have absolute fluxes, which will be the
case with high precision parallax missions such as GAIA.  However, the
more distant objects will have lower precision parallaxes and hence
errors in the {\em mean} flux level.  But even if we completely ignore
distance information (and flux normalise the spectra), the shape of
the spectrum is still a strong indicator of \teff: For example,
Bailer-Jones et al.\ (\cite{bailerjones_98a}) obtained an MK spectral
type precision of 0.8 subtypes ($\Delta\log$\teff=0.010--0.015) across
a wide range of spectral types (B2--M7) using flux normalised
spectra. This is similar to what can be achieved from broad band
photometry, implying that \teff\ determination only requires very low
resolution.

The good performance of `high' resolution spectroscopy (25\,\AA) at
very low SNR ($\sqrt 5$ per pixel) was not expected. It seems to imply
that for a given amount of integration time it may be better to
sacrifice SNR for resolution. It is noteworthy that while the filters
provide good \teff, their ability to determine \met\ and especially
\logg\ is very limited at low SNR.

How do these results compare with classical parametrization methods?
Gray (\cite{gray_92a}) compiles results showing that with photometric
errors below 0.01 magnitudes, the B$-$V colour calibrates \teff\ to
2--3\% (4\% for hotter stars) in the absence of reddening. Slightly
better precision can be obtained from the slope of the Paschen
continuum and size of the Balmer discontinuity.  The latter may also
be used to measure \logg\ to $\pm 0.2$ dex.  With spectra at a few
\AA\ resolution over a similar wavelength range to that used here,
Cacciari et al.\ (\cite{cacciari_87a}) obtained uncertainties in
log~\teff\ and \logg\ of 0.01 and 0.04 respectively.  Sinnerstad
(\cite {sinnerstad_80a}) made uvby,$\beta$ photometric measurements of
B stars, and for uncertainties of 0.005 in $\beta$ and of 0.01 in
u$-$b (i.e.\ SNR $\sim$ 200), infers errors in log~\teff\ and \logg\
of 0.004 and 0.08 respectively. These are similar to or slightly
better than the results for similar stars in
Tables~\ref{restab1}--\ref{restab3} ($\epsilon_3$) at the highest
resolutions. High resolution ($r \leq 0.1$\,\AA)\ spectra have
generally been used to determine metallicity, and in a review, Cayrel
de Strobel (\cite{cayreldestrobel_85a}) notes that metallicity can be
determined to $\pm 0.07$ dex at SNR=250 (but only $\pm 0.2$ dex at
SNR=50) provided the effective temperature and gravity are
approximately known.  At lower SNR (10--20), Jones et al.\
(\cite{jones_96a}) could determine [Fe/H] to $\pm 0.2$ dex for G stars
using a set of spectroscopic indices measured at 1\,\AA\ resolution in
the range 4000--5000\,\AA, again using a known effective temperature.

More recently, Katz et al.\ (\cite{katz_98a}) have used a minimum
distance method to parametrize spectra by finding the closest matching
template spectrum.  The template grid consisted of 211 flux calibrated
spectra (3900--6800\,\AA, $r \simeq 0.1$\,\AA) with
4000\,K\,$\leq$\,\teff\,$\leq$\,6300\,K, $-0.29 \leq$\,[Fe/H]\,$\leq
+0.35$, and \logg\ for dwarfs and giants.  The {\em internal accuracy}
of the method for log~\teff, \logg\ and \met\ was 0.008, 0.28 dex, and
0.16 dex respectively at SNR=100, and 0.009, 0.29 dex and 0.17 dex at
SNR=10. As expected, their results for \logg\ are much better than
those in this paper at the similar temperature range ($\epsilon_1$ in
Table~\ref{restab2}), presumably due to their much higher
resolution. In contrast, their performance for \met\ is similar and
for \teff\ somewhat worse than that in this paper at 500 times lower
resolution. Their results also confirm that at high resolution a lower
SNR leads to very little loss in performance.  Snider et al.\
(\cite{snider_00a}) trained and tested neural networks on a set of 182
real F,G and K spectra over the range 3630--4890\,\AA\ at intermediate
resolution ($\sim$1\AA), and achieved 1$\sigma$ errors in log~\teff,
\logg\ and \met\ of 0.015, 0.41 dex and 0.22 dex respectively, based
on training and testing a network with a set of 182 real F,G and K
spectra.

When judging the relative values of the different resolution/SNR
combinations in this paper, we must also take account of their
implementation `costs', specifically the relative integration times
required.  Usually for a survey, a fixed total amount of integration
time is available for all filters/spectra. In the case of GAIA --
which is continuously rotating -- a star moves across a focal plane
covered with a mosaic of CCDs which are clocked at the rotation rate.
The different filters are fixed to different CCDs, so that as a star
moves across the mosaic it is recorded in different wavelength ranges.
Thus less numerous and/or broader filters would achieve a higher SNR
than more or narrower filters. Some filters could be replaced with a
slitless spectrograph (e.g.\ a prism or grism).  This disperses every
point on the sky and thus gives the full integration time for all
wavelengths, but at the expense of increased sky noise and object
confusion.  These could be reduced by using one or more dichroics to
redirect the light to two or more focal planes.  (Confusion would be
reduced further with GAIA by the fact that each area of sky is
observed at many different position angles over the mission life.)  An
alternative approach is a set of many medium band filters ($\sim 100$
for $r$=100\,\AA\ over the complete wavelength range, although
omission of some filters could be achieved). While this avoids the two
principal disadvantages of the slitless spectrograph, the integration
time per wavelength interval is dramatically reduced.

\section{Development of a survey parametrization system}\label{system}

The development of a complete parametrization system will require
further research, much of which needs to be directed at taking better
account of the true nature of the observational data. Directions and
suggestions for the course of this work are now given.

\subsection{Object selection}

Essentially all of the work in the literature on automated
classification deals with preselected objects. In contrast, an
unpointed survey will pick up a whole range of objects, necessitating
a filtering system to select the stars. Such a system could make use
of both object morphology and spectral features, and systems based on
neural networks (e.g.\ Odewahn et al.\ \cite{odewahn_93a}; Miller \&
Coe \cite{miller_96a}; Serra-Ricart et al.\ \cite{serraricart_96a})
and Principal Components Analysis (Bailer-Jones et al.\
\cite{bailerjones_98a}) have been proposed. Such a system must be
relatively robust and always allow for `unknown' objects which can be
dealt with manually.

\subsection{Model training}

It will be necessary to have a stellar database for training which
takes better account of the larger range of variation present in the
Galactic stellar population. Ideally, a large set of real spectra
across a wide range of physical parameters should be obtained for this
purpose. Good atmospheric models and synthetic spectra are nonetheless
still required for determining their physical parameters and thus for
training the network.  There are two possible approaches to training.
The first is to train on synthetic spectra suitably preprocessed to be
in the same form as the observed spectra (e.g.\ Bailer-Jones et al.\
\cite{bailerjones_97a}). The alternative is to obtain a representative
sample of real spectra with the survey system, calibrate them, and
then use them to train a network. In theory the latter method gives a
better sampling of the true cosmic variance in the spectra, but of
course requires that a representative sample is selected from the
survey data. This sample could be improved as the survey progressed.
Neural networks are fast to train and apply, so it is realistic to
expect that even for a database of $10^9$ objects the network could be
retrained and applied to the whole database in less than a day.

\subsection{Improved stellar models}

More advanced model atmospheres are required for a number of reasons:
\begin{enumerate}
\item{\teff, \met\ and \logg\ do not uniquely
describe a true spectrum. Models sensitive to different
abundance ratios and which include chromospheres (for example) are necessary.}
\item{Kurucz models assume LTE which is known to break down in a
number of regimes (e.g.\ for very hot stars).}
\item{Both the atmospheric models and the line lists lack water
opacity, known to be important for cool stars, thus setting the
current lower \teff\ limit of about 4000\,K.}
\item{Yet more advanced models (which include dust) are required for
very cool stars (L and T dwarfs) and brown dwarfs, of which many will
be found by GAIA.}
\end{enumerate}

\subsection{Reddening}

Of particular importance is insterstellar extinction (reddening),
especially in deep surveys.  The extinction can, in theory, be
determined by the network by training it on artificially reddened
synthetic spectra and providing the network with a ``reddening''
output parameter (or parameters).  This has been demonstrated on
limited data sets by Weaver \& Torres-Dodgen \cite{weaver_95a} and
Gulati, Gupta \& Singh \cite{gulati_97a}, who determined E(B$-$V) to
within 0.05 and 0.08 magnitudes respectively. The latter made use of
6\,\AA\ resolution UV spectra (4850--5380\,\AA). The former used red
spectra (5800--8900\,\AA) at 15\,\AA\ resolution and found that the
spectral type and luminosity class classifications did not degrade
much as reddening was added. It is therefore to be expected that the
parametrizations in this paper will be robust to reddening,
particularly as the spectra have a much larger wavelength coverage.
The filter systems proposed for GAIA were of course designed with
interstellar extinction in mind, and a study of its impact has been
carried out (ESA, in preparation). This work shows that suitable Q
parameters (non-linear combinations of the filter fluxes) used to
determine the physical parameters are largely insensitive to
reddening.  It also claims that narrow band filters are not necessary
for overcoming reddening.  In some parts of the parameter space,
reddening is more problematic (e.g.\ for K stars), largely due to a
degeneracy between it and \teff\ and \logg. However, at intermediate
and high Galactic latitudes it is expected that E(B$-$V) can be
determined to within 0.002 magnitudes.  Munari (\cite{munari_99a})
similarly shows that reddening-free indices exist for the Asiago
filter system. As a neural network also forms non-linear combinations
of the filter fluxes, it is reasonable to suppose that it too will be
robust to redenning, although this will be the subject of future work.

\subsection{Binary systems}

The parametrization model used in a real survey must confront the fact
that most stars are in spatially unresolved multiple systems.
Independent measurement of the physical properties of each component
is desirable and in principle achievable -- when the brightness ratio
is large enough -- by training the network with composite spectra. In
this case the network model would need to have multiple sets of
outputs to deal with each component. An alternative approach is to use
`probabilistic outputs' in which the single output for, say, \teff, is
replaced with a series of outputs in which each value of \teff\ (6000,
6250, 6500 etc.)\ is represented separately.  The network then
evaluates the probability that each temperature is present in the
input spectrum. This method is not recommended, however, as it
eliminates the intrinsically continuous nature of the physical
parameters. It would also greatly increase the number of outputs and
hence the number of free parameters (weights) in the network.

\subsection{Incomplete data}

Object confusion should not result in any overlapped spectrum being
rejected entirely. Rather, it would be better to have a
parametrization model which is robust to missing data. This is a major
challenge for the feedforward network models used in this and most
other papers on automated classification, and will presumably require
some transformation of the input spectrum.  An analysis of the effect
of wavelength coverage on the parameter determination accuracy is
important because a smaller spectral coverage (or {\em coverages} --
it need not be contiguous) would also reduce this confusion.

Finally, the model should make use of all available data. In the case
of GAIA, this means including the data from the high resolution
spectrograph (8470--8700\,\AA\ at 0.75\,\aapix)\ used to measure
radial velocities.  As the inputs to the network need not be
homogenous, there should be no problem incorporating different types
of data.

\section*{Acknowledgements}
I would like to thank Fabio Favata, Gerry Gilmore and Michael Perryman
for useful discussions on this work, in particular within the
context of the GAIA mission.  I am also grateful to Robert Kurucz for
use of his model atmospheres and compiled line lists, and to Richard
Gray for the use of his synthetic spectra generation program.

\end{document}